# Another look at the problem of gauge invariance in quantum field theory

by

Dan Solomon

Rauland-Borg Corporation

3450 W. Oakton Street

Skokie, IL 60076

Phone: 847-324-8337

Email: **dan.solomon@rauland.com**

**Abstract**

It is generally assumed that quantum field theory is gauge invariant. However it is well known that non-gauge invariant terms appear in various calculations. This problem was examined in Refs. [3] and [4] and it was shown that at the formal level quantum field theory in the Schrödinger picture is not, in fact, gauge invariant. It was determined that this problem was due to a mathematical inconsistency relating to the way the vacuum state is defined. In order to shed further light on this problem we consider a "simple" field theory in 1-1D space-time consisting of a quantized fermion field with zero mass in the presence of a classical electromagnetic potential. It can easily be shown that for this situation the equations of motion can be solved exactly in both the Heisenberg and Schrödinger pictures. This allows us to easily identify the source of the mathematical inconsistency.

## 1. Introduction

It is well know that quantum field theory contains anomalies. An anomaly occurs when the result of a calculation does not agree with some underlying symmetry of the theory. Such is the case with gauge invariance.

Quantum field theory is assumed to be gauge invariant [1][2]. A change in the gauge is a change in the electromagnetic potential that does not produce a change in the electromagnetic field. The electromagnetic field is given by,

$$\vec{E} = -\left(\frac{\partial \vec{A}}{\partial t} + \vec{\nabla} A_0\right); \quad \vec{B} = \vec{\nabla} \times \vec{A} \tag{1.1}$$

where $\vec{E}$ is the electric field, $\vec{B}$ is the magnetic field, and $\left(A_0, \vec{A}\right)$ is the electromagnetic potential. A change in the electromagnetic potential that does not produce a change the electromagnetic field is given by,

$$\vec{A} \to \vec{A}' = \vec{A} - \vec{\nabla}\chi, \quad A_0 \to A_0' = A_0 + \frac{\partial \chi}{\partial t} \tag{1.2}$$

where $\chi\left(\vec{x}, t\right)$ is an arbitrary real valued function. Using relativistic notation this can also be written as,

$$A_\nu \to A_\nu' = A_\nu + \partial_\nu \chi \tag{1.3}$$

In order for quantum field theory to be gauge invariant a change in the gauge cannot produce a change in any physical observable such as the current and charge expectation values. However, it is well known that when certain quantities are calculated using standard perturbation theory the results are not gauge invariant [3][4].

For example, consider a system that is initially in the vacuum state. The application of an electromagnetic field will perturb this initial state. The current that is induced in the vacuum state due to the application of the electromagnetic field is called the vacuum current $J_{vac}^{\mu}\left(x\right)$. Using perturbation theory the lowest order term of the vacuum current can be shown to be given by,

$$J_{vac}^{\mu}\left(x\right) = \int \pi^{\mu\nu}\left(x - x'\right) A_\nu\left(x'\right) d^4x' \tag{1.4}$$

where $\pi^{\mu\nu}$ is called the polarization tensor and summation over repeated indices is assumed. The above relationship is normally written in terms of Fourier transformed quantities as,

$$J^{\mu}_{vac}(k) = \pi^{\mu\nu}(k) A_{\nu}(k) \tag{1.5}$$

where k is the 4-momentum. In this case, using relativistic notation, a gauge transformation takes the following form,

$$A_{\nu}(k) \to A'_{\nu}(k) = A_{\nu}(k) + ik_{\nu}\chi(k) \tag{1.6}$$

The change in the vacuum current, $\delta_g J^{\mu}_{vac}(k)$, due to a gauge transformation can be obtained by using (1.6) in (1.5) to yield,

$$\delta_g J^{\mu}_{vac}(k) = ik_{\nu}\pi^{\mu\nu}(k)\chi(k) \tag{1.7}$$

Now the vacuum current is an observable quantity therefore, if quantum theory is gauge invariant, the vacuum current must not be affected by a gauge transformation. Therefore $\delta_g J^{\mu}_{vac}(k)$ must be zero. For this to be true we must have that,

$$k_{\nu}\pi^{\mu\nu}(k) = 0 \tag{1.8}$$

However, when the polarization tensor is calculated it is found that the above relationship does not hold.

For example a calculation of the polarization tensor by [5] yields,

$$\pi^{\mu\nu}(k) = \pi^{\mu\nu}_{G}(k) + \pi^{\mu\nu}_{NG}(k) \tag{1.9}$$

The first term on the right hand side is given by,

$$\pi^{\mu\nu}_{G}(k) = \left(\frac{2q^2}{3\pi}\right)\left(k^{\mu}k^{\nu} - g^{\mu\nu}k^2\right)\int_{2m}^{\infty} dz \frac{\left(z^2 + 2m^2\right)\sqrt{\left(z^2 - 4m^2\right)}}{z^2\left(z^2 - k^2\right)} \tag{1.10}$$

where m is the mass of the electron, q is the electric charge, and $\hbar = c = 1$. The second term on the right of (1.9) is

$$\pi^{\mu\nu}_{NG}(k) = \left(\frac{2q^2}{3\pi}\right) g^{\mu}_{\nu}\left(1 - g^{\mu 0}\right)\int_{2m}^{\infty} dz \frac{\left(z^2 + 2m^2\right)\sqrt{\left(z^2 - 4m^2\right)}}{z^2} \tag{1.11}$$

where there is no summation over the two $\mu$ superscripts that appear on the right. In the above expressions $\pi^{\mu\nu}_{G}(k)$ is gauge invariant because it satisfies $k_{\mu}\pi^{\mu\nu}_{G}(k) = 0$ however $\pi_{NG}^{\mu\nu}(k)$ is not gauge invariant because $k_{\mu}\pi_{G}^{\mu\nu}(k) \neq 0$. Therefore to get a physically valid result it is necessary to "correct" equation (1.9) by dropping $\pi_{NG}^{\mu\nu}$ from the solution. A number of other examples from the literature are considered in Ref. [3]. In all cases the expression for the polarization tensor is of the form of (1.9) and contains the non-gauge invariant term $\pi_{NG}^{\mu\nu}(k)$. In order to obtain a physically correct result this non-gauge invariant term must be removed from the solution. This can be done using the process of "regularization" (see Ref [3] and references therein). While this seems to cure the problem by removing the unwanted terms it introduces an additional step that is not in the original formulation of the theory. The obvious question to ask, then, is why the additional step of regularization is needed in the first place. If the theory is gauge invariant then why does a calculation of the polarization tensor produce non-gauge invariant terms?

This question was originally discussed in some detail in [4] where the problem of gauge invariance was examined for a field theory in the Schrödinger picture consisting of a quantized fermion field in the presence of an unquantized classical electromagnetic field. It is shown in [4] that, when perturbation theory is used, the change in the vacuum current due a gauge transformation is given by,

$$\delta_g J_{vac}(\vec{x},t) = -i\int \chi(\vec{y},t)\langle 0|\left[\hat{\rho}(\vec{y}),\hat{\vec{J}}(\vec{x})\right]|0\rangle d\vec{y} \tag{1.12}$$

In the above expression $\hat{\rho}$ is the charge operator, $\hat{\vec{J}}$ is the current operator, $|0\rangle$ is the vacuum state, and $\chi$ is an arbitrary function which is related to the gauge transformation as specified in (1.2). The quantity $\langle 0|\left[\hat{\rho}(\vec{y}),\hat{J}(\vec{x})\right]|0\rangle$ is commonly called the Schwinger term. If quantum field theory is gauge invariant then $\delta_g J_{vac}(\vec{x},t)$ should be zero due to the fact that $\delta_g J_{vac}(\vec{x},t)$ is the change in the vacuum current due to the gauge transformation. Therefore $\langle 0|\left[\hat{\rho}(\vec{y}),\hat{J}(\vec{x})\right]|0\rangle$ must be zero because $\chi(\vec{y},t)$ is an arbitrary function. However, as was discussed in [3] and [4], it was shown by Schwinger[6] that $\langle 0|\left[\hat{\rho}(\vec{y}),\hat{J}(\vec{x})\right]|0\rangle$ cannot be zero due to the fact that the vacuum state $|0\rangle$ is the state with the lowest free field energy, where the free field energy is defined to be the energy of the quantum state when the electromagnetic potential is set equal to zero. The fact that the Schwinger term is non-zero destroys the gauge invariance of the theory and leads to the presence of non-gauge invariant terms. This was confirmed in [3] where the quantity $\delta_g J_{vac}$ was explicitly calculated in 1-1D space-time and shown to be non-zero.

In Ref. [3] the mathematical consistency of quantum field theory was analyzed. Four elements that are normally considered to be part of quantum field theory were examined. These were; (1) that the Schrödinger equation governs the dynamics of the theory with the Hamiltonian specified by Eq. (2.2) of [3]; (2) the theory is gauge invariant; (3) there is local charge conservation, i.e., the continuity equation is true and; (4) there is lower bound to the free field energy. It was shown that these elements of QFT are not mathematically consistent. Specifically item (2) is incompatible with item (4), that is, if QFT is gauge invariant then there cannot be a lower bound to the free field energy. However it can be readily shown that the vacuum state is a lower bound to the free field energy. Therefore, as discussed in [3] and [4], quantum field theory in the Schrödinger picture is not gauge invariant at the formal level. This, then, explains why non-gauge invariant terms appear in the polarization tensor. Since the theory in not gauge invariant in the first place it would be expected that the results of calculations are also not gauge invariant. This is, of course, exactly what happens.

It is the purpose of this paper to continue the discussion started in Refs. [3] and [4]. We will examine a simple field theory in 1-1D space-time consisting of a quantized fermion field with zero mass in the presence of a classical non-quantized electric field. The advantage of working with this simple field theory is that the equations of motion can be solved for exactly in both the Heisenberg and Schrödinger pictures for an arbitrary electromagnetic field. It is then easy to show that inconsistencies will arise when the vacuum state is defined in the conventional manner. Of course the field theory to be analyzed here is considerably simplified from the "standard model". The reason for studying it is that is can be fully analyzed without the need for perturbation theory. The hope is that the understanding that we gain from the studying the simple theory can be applied to resolving some of the problems of the more complicated theory.

The discussion will proceed as follows. In Section 2 the different approaches leading to the Heisenberg and Schrödinger pictures will be discussed. In Section 3 the Heisenberg picture is developed and a solution is obtained for the time dependent Heisenberg field operator. From this it is easy to show that the Heisenberg picture is gauge invariant. In addition we show that there is no lower bound to the free field energy in the Heisenberg picture. In Section 4 the Schrödinger picture is developed. We obtain a solution for the time dependent Schrödinger state vector and show that the two pictures are equivalent. This means that the Schrödinger picture is gauge invariant and that there is no lower bound to the free field energy in the Schrödinger picture. In Section 5 we introduce the vacuum state $|0\rangle$ according to normal procedures. It is shown that the state $|0\rangle$ is a lower bound to the free field energy. This results in a mathematical inconsistency since we have just shown that there is no lower bound to the free field energy in both the Schrödinger and Heisenberg pictures. In Section 6 we discuss the results achieved up to this point. Next, in Section 7, we examine the effect this inconsistency has on the problem of gauge invariance by examining the Schwinger term. Based on previous work and as discussed above we know that in order for the theory to be gauge invariant the Schwinger term must be zero. The Schwinger term is calculated using $|0\rangle$ as the vacuum state and is shown to be non-zero. Then in Section 8 we redefine that vacuum state so that it is no longer the lower bound to the free field energy. We call this state $|0_R\rangle$. We then recalculate the Schwinger term using $|0_R\rangle$ as the vacuum state and show that in this case the Schwinger term is zero. This, of course, is what is required for a gauge invariant theory. The results are then summarized in Section 9.

## 2. Heisenberg versus Schrödinger picture.

In quantum field theory a quantum system, at a given point in time, is specified by the state vector $|\Omega\rangle$ and field operator $\hat{\psi}(z)$ where $z$ is the space coordinate in 1-1D space-time. We will write this as the pair $(|\Omega\rangle, \hat{\psi})$. Let the state vector $|\Omega\rangle$ and the field operator $\hat{\psi}(z)$ be defined at some initial point in time, say, $t = 0$. This may be taken as the initial conditions of the quantum system. Now there are two ways to handle the time evolution of the system. In the Schrödinger picture it is assumed that field operator $\hat{\psi}(z)$ is constant in time and the time dependence of the system goes with the state vector $|\Omega(t)\rangle$. In the Heisenberg picture the time dependence is assigned to the field operator $\hat{\psi}(z,t)$ and the state vector $|\Omega\rangle$ remains constant in time.

Note that at the initial time, $t=0$, both pictures are identical. We will assume that at the initial time $t=0$ the normalized state vector is represented by $|\Omega(0)\rangle$ and that the field operator is in some initial unperturbed state which is designated by $\hat{\psi}_0(z)$ and will be specified later. The system at this initial time is represented by $(|\Omega(0)\rangle, \hat{\psi}_0(z))$. In the Heisenberg picture this state evolves into the system $(|\Omega(0)\rangle, \hat{\psi}(z,t))$ where $\hat{\psi}(z,0)=\hat{\psi}_0(z)$. In the Schrödinger picture this evolves into the system $(|\Omega(t)\rangle, \hat{\psi}_0(z))$. It will be shown that the expectation value of operators are equivalent in both pictures. Because of this both pictures are considered to be equivalent representations of quantum field theory.

Note in order to avoid confusion we will often use the subscript $h$ to indicate that a quantity is associated with the Heisenberg picture and the subscript $S$ to indicate a quantity associated with the Schrödinger picture. In a few cases the subscript is not used in order to simplify the notation however it should be obvious from the definition which picture we are working in.

**3. The Heisenberg picture and gauge invariance.**

Now consider a "simple" field theory in 1-1D space-time consisting of a quantized fermion field with non-interacting zero mass fermions in the presence of a classical electromagnetic field. In this case the time evolution of the field operator in the Heisenberg picture is given by,

$$i\frac{\partial\hat{\psi}(z,t)}{\partial t} = H_D(A_0, A_1)\hat{\psi}(z,t) \tag{3.1}$$

where,

$$H_D(A_0, A_1) = H_0 - q\sigma_3 A_1 + qA_0 \tag{3.2}$$

and,

$$H_0 = -i\sigma_3 \frac{\partial}{\partial z} \tag{3.3}$$

In the above expression the field operator $\hat{\psi}(z,t)$ is a two-spinor and the electromagnetic potential $(A_0, A_1)$ is taken to be a classical, unquantized, real valued quantity. Also $q$ is the electric charge and $\sigma_3$ is the Pauli matrix with,

$$\sigma_3 = \begin{pmatrix} 1 & 0 \\ 0 & -1 \end{pmatrix} \tag{3.4}$$

Note also that in the above equations we use $\hbar = c = 1$. Furthermore, note that in 1-1D space-time there is no magnetic field and the electric field is given by,

$$E = -\left( \frac{\partial A_1}{\partial t} + \frac{\partial A_0}{\partial z} \right) \tag{3.5}$$

The solution to (3.1) is given by,

$$\hat{\psi}(z,t) = W(z,t)\hat{\psi}_0(z,t) \tag{3.6}$$

where $\hat{\psi}_0(z,t)$ is the solution to the free field equation,

$$i\frac{\partial \hat{\psi}_0(z,t)}{\partial t} = H_0 \hat{\psi}_0(z,t) \tag{3.7}$$

and the quantity $W(z,t)$ is given by,

$$W(z,t) = \begin{pmatrix} e^{-ic_1} & 0 \\ 0 & e^{-ic_2} \end{pmatrix} \tag{3.8}$$

where $c_1$ and $c_2$ satisfy the following differential equations,

$$\frac{\partial c_1}{\partial t} + \frac{\partial c_1}{\partial z} = qA_0 - qA_1 \tag{3.9}$$

and,

$$\frac{\partial c_2}{\partial t} - \frac{\partial c_2}{\partial z} = qA_0 + qA_1 \tag{3.10}$$

In the Heisenberg picture the current expectation value for a normalized state vector $|\Omega(0)\rangle$ is given by,

$$J_{h,e}(z,t) = \langle \Omega(0) | \hat{J}_h(z,t) | \Omega(0) \rangle \tag{3.11}$$

where the current operator $\hat{J}_h(z,t)$ is defined,

$$\hat{J}_h(z,t) = q\hat{\psi}^{\dagger}(z,t)\sigma_3\hat{\psi}(z,t) \tag{3.12}$$

Similarly the Heisenberg picture charge expectation value is defined by

$$\rho_{h,e}(z,t) = \langle\Omega(0)|\hat{\rho}_h(z,t)|\Omega(0)\rangle \tag{3.13}$$

where the charge operator is,

$$\hat{\rho}_h(z,t) = q\hat{\psi}^{\dagger}(z,t)\hat{\psi}(z,t) \tag{3.14}$$

Recall that the $h$ in the subscript indicates that we are working in the Heisenberg picture. Using the above definitions along with (3.1) we can readily show that,

$$\frac{\partial J_{h,e}}{\partial t} + \frac{\partial \rho_{h,e}}{\partial z} = 0\,; \qquad \frac{\partial \rho_{h,e}}{\partial t} + \frac{\partial J_{h,e}}{\partial z} = 0 \tag{3.15}$$

Using (3.6) in (3.12) we obtain,

$$\hat{J}_h(z,t) = q\hat{\psi}_0(z,t)W^{\dagger}(z,t)\sigma_3 W(z,t)\hat{\psi}_0(z,t) \tag{3.16}$$

It is easy to show that $W^{\dagger}(z,t)\sigma_3 W(z,t) = \sigma_3$ therefore we obtain,

$$\hat{J}_h(z,t) = q\hat{\psi}_0^{\dagger}(z,t)\sigma_3\hat{\psi}_0(z,t) \tag{3.17}$$

Therefore the current expectation value in the Heisenberg picture is,

$$J_{h,e}(z,t) = q\langle\Omega(0)|\hat{\psi}_0^{\dagger}(z,t)\sigma_3\hat{\psi}_0(z,t)|\Omega(0)\rangle \tag{3.18}$$

By a similar argument we obtain,

$$\rho_{h,e}(z,t) = q\langle\Omega(0)|\hat{\psi}_0^{\dagger}(z,t)\hat{\psi}_0(z,t)|\Omega(0)\rangle \tag{3.19}$$

Recall that $\hat{\psi}_0(z,t)$ is a solution to (3.7) and is independent of the electromagnetic potential. Therefore the current expectation value $J_{h,e}(z,t)$ is also independent of the electromagnetic potential. The same reasoning applies to the charge expectation value.

Now we can address the question of whether the theory is gauge invariant. Since the current and charge expectation values are independent of the electromagnetic potential the theory is obviously gauge invariant because a change in the gauge will not effect the current and charge expectation values.

Next define the free field energy of a normalized state vector $\left|\Omega(0)\right\rangle$ in the Heisenberg picture by,

$$\xi_{0h}(t)=\left\langle\Omega(0)\right|\hat{H}_{0h}(t)\left|\Omega(0)\right\rangle \tag{3.20}$$

where,

$$\hat{H}_{0h}(t)=\int\hat{\psi}^{\dagger}(z,t)H_0\hat{\psi}(z,t)dz \tag{3.21}$$

The question that we want to address at this point is whether or not there is a lower bound to the free field energy $\xi_{0h}(t)$. To determine this start with the system at the initial time $t=0$ in the initial state $\left(\left|\Omega(0)\right\rangle,\hat{\psi}_0(z)\right)$. Now apply an electromagnetic potential where $A_1=0$ and $A_0$ is non-zero. For $t>0$ the quantum state is given by $\left(\left|\Omega(0)\right\rangle,\hat{\psi}(z,t)\right)$ where $\hat{\psi}(z,t)$ is given by (3.6). Use (3.6) and (3.21) in (3.20) to obtain,

$$\xi_{0h}(t)=\left\langle\Omega(0)\right|\int\hat{\psi}_0^{\dagger}(z,t)W^{\dagger}H_0W\hat{\psi}_0(z,t)dz\left|\Omega(0)\right\rangle \tag{3.22}$$

Use (3.3) and (3.8) in the above to obtain,

$$\xi_{0h}(t)=\left\langle\Omega(0)\right|\int\hat{\psi}_0^{\dagger}(z,t)\begin{pmatrix}-\partial c_1/\partial z & 0\\ 0 & \partial c_2/\partial z\end{pmatrix}\hat{\psi}_0(z,t)dz\left|\Omega(0)\right\rangle \tag{3.23}$$

Using (3.18) and (3.21) we obtain,

$$\xi_{0h}(t)=-\frac{1}{2q}\int\left(J_{h,e}\frac{\partial(c_1+c_2)}{\partial z}+\rho_{h,e}\frac{\partial(c_1-c_2)}{\partial z}\right)dz \tag{3.24}$$

Take the derivative with respect to time of the above and use (3.15) to obtain,

$$\frac{\partial\xi_{0h}(t)}{\partial t}=-\frac{1}{2q}\int\left(-\frac{\partial\rho_{h,e}}{\partial z}\frac{\partial(c_1+c_2)}{\partial z}+J_{h,e}\frac{\partial^2(c_1+c_2)}{\partial t\partial z}-\frac{\partial J_{h,e}}{\partial z}\frac{\partial(c_1-c_2)}{\partial z}+\rho_{h,e}\frac{\partial^2(c_1-c_2)}{\partial t\partial z}\right)dz \tag{3.25}$$

Assume reasonable boundary conditions and integrate by parts to obtain,

$$\frac{\partial\xi_{0h}(t)}{\partial t}=-\frac{1}{2q}\int\left(\rho_{h,e}\frac{\partial}{\partial z}\left(\frac{\partial(c_1-c_2)}{\partial t}+\frac{\partial(c_1+c_2)}{\partial z}\right)+J_{h,e}\frac{\partial}{\partial z}\left(\frac{\partial(c_1+c_2)}{\partial t}+\frac{\partial(c_1-c_2)}{\partial z}\right)\right)dz \tag{3.26}$$

Use (3.9) and (3.10) along with the previous assumption that $A_1=0$ and (3.5) to obtain,

$$\frac{\partial \xi_{0h}(t)}{\partial t} = \int dz J_{h,e}(z,t) E(z,t) \tag{3.27}$$

Therefore the free field energy at time $t_f$ is,

$$\xi_{0h}(t_f) = \xi_{0h}(0) + \int_0^{t_f} dt \int dz J_{h,e}(z,t) E(z,t) \tag{3.28}$$

Now, based on the previous discussion, the current expectation value $J_{h,e}(z,t)$ is independent of the electromagnetic potential. Therefore the quantities $J_{h,e}(z,t)$ and $E(z,t)$ are independent of each other. Assume that we have chosen the state vector $|\Omega(0)\rangle$ so that the current expectation value $J_{h,e}(z,t)$ is non-zero. It will be shown later that it is possible to do this. In this case we can always find a $E(z,t)$ so that there is no lower bound to $\xi_{0h}(t_f)$. For example, let $E(z,t) = -fJ_{h,e}(z,t)$ where $f$ is a constant. Then (3.28) becomes,

$$\xi_{0h}(t_f) = \xi_{0h}(0) - f\int_0^{t_f} dt \int dz \left(J_{h,e}(z,t)\right)^2 \tag{3.29}$$

For a non-zero $J_{h,e}(z,t)$ the integral is greater than zero. Therefore as $f \to \infty$ we have that $\xi_{0h}(t_f) \to -\infty$. Therefore that is no lower bound to the free field energy in the Heisenberg picture.

**4. The Schrödinger picture.**

In this section we will set up the theory in the Schrödinger picture. In this case the field operator $\hat{\psi}_0(z)$ is constant in time and time evolution of the system is reflected in the state vector $|\Omega(t)\rangle$. Assume that the initial unperturbed field operator $\hat{\psi}_0(z)$ obeys the anticommutation relationships,

$$\hat{\psi}_{0,\alpha}(z)\hat{\psi}^\dagger_{0,\beta}(z') + \hat{\psi}^\dagger_{0,\beta}(z')\hat{\psi}_{0,\alpha}(z) = \delta_{\alpha\beta}\delta(z-z') \tag{4.1}$$

where $\alpha$ and $\beta$ are spinor indices and all other anticommutations are zero. In the Schrödinger picture the state vector evolves in time according to the equation,

$$i\frac{\partial|\Omega(t)\rangle}{\partial t} = \hat{H}|\Omega(t)\rangle \text{ and } -i\frac{\partial\langle\Omega(t)|}{\partial t} = \langle\Omega(t)|\hat{H} \tag{4.2}$$

where,

$$\hat{H} = \int \hat{\psi}_0^\dagger(z) H_D(A_0, A_1) \hat{\psi}_0(z)\,dz \tag{4.3}$$

We can use (4.1) to show that the solution to (4.2) is,

$$|\Omega(t)\rangle = e^{-i\hat{G}_0(t)} e^{-i\hat{H}_0 t} |\Omega(0)\rangle \tag{4.4}$$

where,

$$\hat{H}_0 = \int \hat{\psi}_0^\dagger(z) H_0 \hat{\psi}_0(z)\,dz \tag{4.5}$$

and,

$$\hat{G}_0(t) = \int \hat{\psi}_0^\dagger(z) F(z,t) \hat{\psi}_0(z)\,dz \tag{4.6}$$

where,

$$F(z,t) = \begin{pmatrix} c_1(z,t) & 0 \\ 0 & c_2(z,t) \end{pmatrix} \tag{4.7}$$

Recall that $c_1(z,t)$ and $c_2(z,t)$ satisfy equations (3.9) and (3.10), respectively.

We will now prove that (4.4) is the solution to (4.2) as follows. Start by using the anticommutation relationships (4.1) to show that,

$$\left[\hat{\psi}_{0,\alpha}(z), \hat{G}_0\right] = F_{\alpha\beta}(z,t)\hat{\psi}_{0,\beta}(z) = \left\langle \begin{matrix} c_1(z,t)\hat{\psi}_{0,1}(z) \text{ if } \alpha = 1 \\ c_2(z,t)\hat{\psi}_{0,2}(z) \text{ if } \alpha = 2 \end{matrix} \right. \tag{4.8}$$

Next use the Baker-Campell-Hausdorff relationships which state that,

$$e^{+O_A} O_B e^{-O_A} = O_B + [O_A, O_B] + \frac{1}{2}[O_A, [O_A, O_B]] + \ldots \tag{4.9}$$

Using the above along with (4.8) yields,

$$e^{i\hat{G}_0} \hat{\psi}_0 e^{-i\hat{G}_0} = W\hat{\psi}_0 \text{ and } e^{i\hat{G}_0} \hat{\psi}_0^\dagger e^{-i\hat{G}_0} = \hat{\psi}_0^\dagger W^\dagger \tag{4.10}$$

Also it follows from the anticommutation relationships that,

$$\left[\frac{\partial \hat{G}_0}{\partial t}, \hat{G}_0\right] = 0 \tag{4.11}$$

Next use (4.4) in (4.2) along with the above expression to obtain,

$$\left( e^{-i\hat{G}_0} \frac{\partial \hat{G}_0}{\partial t} e^{-i\hat{H}_0 t} + e^{-i\hat{G}_0} e^{-i\hat{H}_0 t} \hat{H}_0 \right) |\Omega(0)\rangle = \hat{H} e^{-i\hat{G}_0} e^{-i\hat{H}_0 t} |\Omega(0)\rangle \tag{4.12}$$

Multiply from the left by $e^{i\hat{G}_0}$ to yield,

$$\left(\frac{\partial \hat{G}_0}{\partial t} e^{-i\hat{H}_0 t} + e^{-i\hat{H}_0 t}\hat{H}_0\right)\left|\Omega(0)\right\rangle = e^{i\hat{G}_0}\hat{H}e^{-i\hat{G}_0}e^{-i\hat{H}_0 t}\left|\Omega(0)\right\rangle \tag{4.13}$$

Next use (4.10) and the fact that $e^{-i\hat{G}_0} H_D e^{+i\hat{G}_0} = H_D$ to obtain the following expression,

$$\begin{aligned} e^{i\hat{G}_0}\hat{H}e^{-i\hat{G}_0} &= \int e^{i\hat{G}_0}\hat{\psi}_0^\dagger(z)e^{-i\hat{G}_0}H_D(A_0,A_1)e^{i\hat{G}_0}\hat{\psi}_0(z)e^{-i\hat{G}_0}dz \\ &= \int \hat{\psi}_0^\dagger(z)W^\dagger H_D(A_0,A_1)W\hat{\psi}_0(z)\,dz \end{aligned} \tag{4.14}$$

We can then show that,

$$W^\dagger H_D(A_0,A_1)W = \begin{pmatrix} -\frac{\partial c_1}{\partial z} - qA_1 + qA_0 & 0 \\ 0 & \frac{\partial c_2}{\partial z} + qA_1 + qA_0 \end{pmatrix} + H_0 \tag{4.15}$$

Use these results in (4.13) to obtain,

$$\frac{\partial \hat{G}_0}{\partial t} e^{-i\hat{H}_0 t}\left|\Omega(0)\right\rangle = \left(\int \hat{\psi}_0^\dagger \begin{pmatrix} -\frac{\partial c_1}{\partial z} - qA_1 + qA_0 & 0 \\ 0 & \frac{\partial c_2}{\partial z} + qA_1 + qA_0 \end{pmatrix}\hat{\psi}_0 dz\right) e^{-i\hat{H}_0 t}\left|\Omega(0)\right\rangle \tag{4.16}$$

Use (4.7) to show that,

$$\frac{\partial \hat{G}_0}{\partial t} = \int \hat{\psi}_0^\dagger \begin{pmatrix} \frac{\partial c_1}{\partial t} & 0 \\ 0 & \frac{\partial c_2}{\partial t} \end{pmatrix}\hat{\psi}_0 dz \tag{4.17}$$

Next use (3.9) and (3.10) along with (4.17) to show that (4.16) is true. This proves that (4.4) is the solution of (4.2).

Next we will consider the question of gauge invariance in the Schrödinger picture. In the Schrödinger picture the current and charge expectation values are given by,

$$J_{S,e}(z,t) = \left\langle\Omega(t)\right|\hat{J}_S(z)\left|\Omega(t)\right\rangle \text{ and } \rho_{S,e}(z,t) = \left\langle\Omega(t)\right|\hat{\rho}_S(z)\left|\Omega(t)\right\rangle \tag{4.18}$$

where the Schrödinger picture current and charge operators are,

$$\hat{J}_S(z) = q\hat{\psi}_0^\dagger(z)\sigma_3\hat{\psi}_0(z) \text{ and } \hat{\rho}_S(z) = q\hat{\psi}_0^\dagger(z)\hat{\psi}_0(z) \tag{4.19}$$

Use (4.4) in the above equations to obtain,

$$J_{S,e}(z,t) = q\left\langle\Omega(0)\right|e^{i\hat{H}_0 t}e^{i\hat{G}_0(t)}\hat{\psi}_0^\dagger(z)\sigma_3\hat{\psi}_0(z)e^{-i\hat{G}_0(t)}e^{-i\hat{H}_0 t}\left|\Omega(0)\right\rangle \tag{4.20}$$

Use (4.10) in the above to obtain,

$$J_{S,e}(z,t)=q\left\langle\Omega(0)\right|e^{i\hat{H}_0 t}\hat{\psi}_0^\dagger(z)W^\dagger\sigma_3 W\hat{\psi}_0(z)e^{-i\hat{H}_0 t}\left|\Omega(0)\right\rangle \tag{4.21}$$

This becomes,

$$J_{S,e}(z,t)=q\left\langle\Omega(0)\right|e^{i\hat{H}_0 t}\hat{\psi}_0^\dagger(z)\sigma_3\hat{\psi}_0(z)e^{-i\hat{H}_0 t}\left|\Omega(0)\right\rangle \tag{4.22}$$

Similarly,

$$\rho_{S,e}(z,t)=q\left\langle\Omega(0)\right|e^{i\hat{H}_0 t}\hat{\psi}_0^\dagger(z)\hat{\psi}_0(z)e^{-i\hat{H}_0 t}\left|\Omega(0)\right\rangle \tag{4.23}$$

Now we see that both the current and charge expectation values are independent of the electric potential. Therefore the Schrödinger picture is gauge invariant.

Now using the above results we will show that expectation values in both pictures are equal. First we will show that,

$$\hat{\psi}_0(z,t)=e^{i\hat{H}_0 t}\hat{\psi}_0(z)e^{-i\hat{H}_0 t} \tag{4.24}$$

To prove this use the derivative of the above expression with respect to time to obtain,

$$i\frac{\partial\hat{\psi}_0(z,t)}{\partial t}=e^{i\hat{H}_0 t}\left[\hat{\psi}_0(z),\hat{H}_0\right]e^{-i\hat{H}_0 t} \tag{4.25}$$

We can use the anti-commutation relationships (4.1) to obtain,

$$\left[\hat{\psi}_0(z),\hat{H}_0\right]=H_0\hat{\psi}_0(z) \tag{4.26}$$

Use this in (4.25) along with (4.24) to obtain,

$$i\frac{\partial\hat{\psi}_0(z,t)}{\partial t}=e^{i\hat{H}_0 t}H_0\hat{\psi}_0(z)e^{-i\hat{H}_0 t}=H_0 e^{i\hat{H}_0 t}\hat{\psi}_0(z)e^{-i\hat{H}_0 t}=H_0\hat{\psi}_0(z,t) \tag{4.27}$$

This is identical to (3.7) which proves that (4.24) is true. Use (4.24) along with $e^{-i\hat{H}_0 t}\sigma_3 e^{i\hat{H}_0 t}=\sigma_3$ in (4.22) to obtain,

$$J_{S,e}(z,t)=q\left\langle\Omega(0)\right|\hat{\psi}_0^\dagger(z,t)\sigma_3\hat{\psi}_0(z,t)\left|\Omega(0)\right\rangle \tag{4.28}$$

Referring to (3.18) we see that $J_{S,e}(z,t)=J_{h,e}(z,t)$. Similarly we have $\rho_{S,e}(z,t)=\rho_{h,e}(z,t)$. We can also show that the free field energy is the same in both pictures. The free field energy in the Schrödinger picture is given by,

$$\xi_{0S}(t)=\left\langle\Omega(t)\right|\hat{H}_0\left|\Omega(t)\right\rangle \tag{4.29}$$

We will start by showing that the free field energy in the Schrödinger picture is equal to the free field energy in the Heisenberg picture. Use (4.4) in the above to obtain,

$$\xi_{0S}(t) = \langle \Omega(0) | e^{i\hat{H}_0 t} e^{i\hat{G}_0} \hat{H}_0 e^{-i\hat{G}_0} e^{-i\hat{H}_0 t} | \Omega(0) \rangle \tag{4.30}$$

Using (4.10) we obtain,

$$e^{i\hat{G}_0} \hat{H}_0 e^{-i\hat{G}_0} = e^{i\hat{G}_0} \int \hat{\psi}_0^\dagger(z) H_0 \hat{\psi}_0(z) dz e^{-i\hat{G}_0} = \int \hat{\psi}_0^\dagger(z) W^\dagger H_0 W \hat{\psi}_0(z) dz \tag{4.31}$$

Use (4.31) and (4.24) to obtain,

$$e^{i\hat{H}_0 t} e^{i\hat{G}_0} \hat{H}_0 e^{-i\hat{G}_0} e^{-i\hat{H}_0 t} = \int e^{i\hat{H}_0 t} \hat{\psi}_0^\dagger(z) W^\dagger H_0 W \hat{\psi}_0(z) e^{-i\hat{H}_0 t} dz = \int \hat{\psi}_0^\dagger(z,t) W^\dagger H_0 W \hat{\psi}_0(z,t) dz \tag{4.32}$$

Use (3.6) in the above to obtain,

$$e^{i\hat{H}_0 t} e^{i\hat{G}_0} \hat{H}_0 e^{-i\hat{G}_0} e^{-i\hat{H}_0 t} = \int \hat{\psi}^\dagger(z,t) H_0 \hat{\psi}(z,t) dz \tag{4.33}$$

Use this result and (3.21) to obtain,

$$\xi_{0S}(t) = \langle \Omega(0) | \hat{H}_{0h}(t) | \Omega(0) \rangle \tag{4.34}$$

Refer to equation (3.20) to show that $\xi_{0S}(t) = \xi_{0h}(t)$. Therefore the free field energy is the same in both picture.

We have shown that expectation values are the same in both pictures. This means that the Heisenberg and Schrödinger pictures are equivalent representations of quantum field theory. Therefore the result that there is no lower bound to the free field energy that we proved is true for the Heisenberg picture also holds true for the Schrödinger picture.

## 5. The field operator and vacuum state.

We will now express the Schrödinger field operator $\hat{\psi}_0(z)$ in terms of the usual destruction and creation operators. This must be done in a way that is consistent with the anti-commutation relationships. We can express $\hat{\psi}_0(z)$ as,

$$\hat{\psi}_0(z) = \hat{\psi}_{0,+1}(z) + \hat{\psi}_{0,-1}(z) \tag{5.1}$$

where,

$$\hat{\psi}_{0,+1}(z) = \sum_{p>0} \left( \hat{b}_{p,+1} \phi_{p,+1}(z) + \hat{d}_{p,+1}^\dagger \phi_{-p,+1}(z) \right) \tag{5.2}$$

and,

$$\hat{\psi}_{0,-1}(z) = \sum_{p<0} \left( \hat{b}_{p,-1} \phi_{p,-1}(z) + \hat{d}_{p,-1}^\dagger \phi_{-p,-1}(z) \right) \tag{5.3}$$

where the $\hat{b}_{p,s}$ ($\hat{b}^{\dagger}_{p,s}$) are the destruction(creation) operators for fermions in the state $\phi_{p,s}(z)$ and $\hat{d}_{p,s}$ ($\hat{d}^{\dagger}_{p,s}$) are the destruction(creation) operators for positrons in the state $\phi_{-p,+1}(z)$ where $s=\pm 1$ is the spin state and $p$ is the momentum.

The destruction and creation operators satisfy the anticommutation relation,

$$\hat{b}_{p,s}\hat{b}^{\dagger}_{q,r}+\hat{b}^{\dagger}_{q,r}\hat{b}_{p,s}=\delta_{pq}\delta_{sr}\,;\;\; \hat{d}_{p,s}\hat{d}^{\dagger}_{q,r}+\hat{d}^{\dagger}_{q,r}\hat{d}_{p,s}=\delta_{pq}\delta_{rs} \tag{5.4}$$

and all other anticommutations are zero. The $\phi_{p,s}(z)$ can be expressed by,

$$\phi_{p,+1}(z)=\frac{1}{\sqrt{L}}\begin{pmatrix}1\\0\end{pmatrix}e^{ipz}\,;\;\phi_{p,-1}(z)=\frac{1}{\sqrt{L}}\begin{pmatrix}0\\1\end{pmatrix}e^{ipz} \tag{5.5}$$

where $L$ is the 1D integration volume. We assume periodic boundary conditions so that the momentum $p_r=2\pi r/L$ where $r$ is an integer. In the following discussion this subscript will be dropped when referring to the momentum.

The $\phi_{p,s}(z)$ are basis state solutions of the free field Dirac equation and satisfy,

$$H_0\phi_{p,s}(z)=sp\phi_{p,s}(z) \tag{5.6}$$

Use the above relationships in (4.5) to obtain,

$$\hat{H}_0=\sum_{p>0}|p|\left(\hat{b}^{\dagger}_{p,1}\hat{b}_{p,1}-\hat{d}_{p,1}\hat{d}^{\dagger}_{p,1}\right)+\sum_{p<0}|p|\left(\hat{b}^{\dagger}_{p,-1}\hat{b}_{p,-1}-\hat{d}_{p,-1}\hat{d}^{\dagger}_{p,-1}\right) \tag{5.7}$$

Now that we have specified the field operator at the initial time we must define the state vectors on which the field operators act. First define the vacuum state $|0\rangle$ as the state that is destroyed by all electron and positron destruction operators, i.e.,

$$\hat{b}_{p,s}|0\rangle=\hat{d}_{p,s}|0\rangle=0 \tag{5.8}$$

This also yields the dual relationships,

$$\langle 0|\hat{b}^{\dagger}_{p,s}=\langle 0|\hat{d}^{\dagger}_{p,s}=0 \tag{5.9}$$

Then, using the above along with (5.7) and we obtain,

$$\hat{H}_0|0\rangle=\varepsilon(|0\rangle)|0\rangle \tag{5.10}$$

where the eigenvalue $\varepsilon(|0\rangle)$ is given by,

$$\varepsilon(|0\rangle)=-\sum_{p}|p| \tag{5.11}$$

This is obviously a divergent quantity. However that will not be a problem because we are actually concerned with differences in the energy and not the actual value. Additional eigenstates $|n\rangle$ are formed by acting on the vacuum state $|0\rangle$ with the various combinations of the creation operators $b_{p,s}^{\dagger}$ and $d_{p,s}^{\dagger}$. The effect of doing this is to create states with positive energy with respect to the vacuum state. The set of eigenstates $|n\rangle$ (which includes the vacuum state $|0\rangle$) form an orthonormal basis that satisfies the following relationships,

$$\hat{H}_0|n\rangle = \varepsilon(|n\rangle)|n\rangle \text{ where } \varepsilon(|n\rangle) > \varepsilon(|0\rangle) \text{ for } |n\rangle \neq |0\rangle \qquad (5.12)$$

and

$$\langle n|m\rangle = \delta_{mn} \qquad (5.13)$$

Any arbitrary normalized state $|\Omega\rangle$ can be expanded in terms of these basis states,

$$|\Omega\rangle = \sum_n c_n |n\rangle \qquad (5.14)$$

where the normalization condition is,

$$\langle\Omega|\Omega\rangle = \sum_n |c_n|^2 = 1 \qquad (5.15)$$

The free field energy expectation value of this state is,

$$\langle\Omega|\hat{H}_0|\Omega\rangle = \sum_n |c_n|^2 \varepsilon(|n\rangle) \qquad (5.16)$$

Use the above to show that,

$$\langle\Omega|\hat{H}_0|\Omega\rangle - \langle 0|\hat{H}_0|0\rangle = \sum_n |c_n|^2 \left(\varepsilon(|n\rangle) - \varepsilon(|0\rangle)\right) \geq 0 \qquad (5.17)$$

This can be also written as,

$$\langle\Omega|\hat{H}_0|\Omega\rangle - \langle 0|\hat{H}_0|0\rangle \geq 0 \text{ for all } |\Omega\rangle \qquad (5.18)$$

The key result of this section is that in the Schrödinger picture there is a lower bound to the free field energy of an arbitrary normalized state vector $|\Omega\rangle$. This lower bound is the quantity $\langle 0|\hat{H}_0|0\rangle$ which is the free field energy of the vacuum state.

**6. Discussion.**

Based on the results of the last section it is evident that there is a mathematical inconsistency in the theory. We have shown in Section 3 that in the Heisenberg picture there is no lower bound to the free field energy if there exists a state with non-zero current. Using the formulation of the last section it is easy to find such a state. For example suppose at the initial time $t=0$ the state vector $\left|\Omega'(0)\right\rangle$ is given by,

$$\left|\Omega'(0)\right\rangle = \frac{1}{\sqrt{2}}\left(\hat{b}^{\dagger}_{p,1} + \hat{b}^{\dagger}_{q,1}\right)|0\rangle \tag{5.19}$$

The free field energy of this state is,

$$\left\langle\Omega'(0)\right|\hat{H}_0\left|\Omega'(0)\right\rangle = \frac{1}{2}\langle 0|\left(\hat{b}_{p,1} + \hat{b}_{q,1}\right)\hat{H}_0\left(\hat{b}^{\dagger}_{p,1} + \hat{b}^{\dagger}_{q,1}\right)|0\rangle = \frac{1}{2}\left(|p|+|q|\right) + \langle 0|\hat{H}_0|0\rangle \tag{5.20}$$

Suppose this state evolves forward in time in the presence of some electric. Referring to (4.22) we can show that the current expectation value of this state is,

$$J_{S,e}(z,t) = \frac{1}{2L}\left(1+\cos\left((p-q)(z-t)\right)\right) \tag{5.21}$$

which is obviously non-zero. Recall that this quantity is independent of the electric field. In order to calculate the free field energy at some time $t_f > 0$ refer to Eq. (3.28) and use the fact that expectation values including the free field energy are the same in both pictures to obtain,

$$\left\langle\Omega'(t_f)\right|\hat{H}_0\left|\Omega'(t_f)\right\rangle = \left\langle\Omega'(0)\right|\hat{H}_0\left|\Omega'(0)\right\rangle + \int_0^{t_f} dt \int dz J_{S,e}(z,t)E(z,t) \tag{5.22}$$

Subtract $\langle 0|\hat{H}_0|0\rangle$ from both sides and use (5.20) to obtain,

$$\left\langle\Omega'(t_f)\right|\hat{H}_0\left|\Omega'(t_f)\right\rangle - \langle 0|\hat{H}_0|0\rangle = \frac{1}{2}\left(|p|+|q|\right) + \int_0^{t_f} dt \int dz J_{S,e}(z,t)E(z,t) \tag{5.23}$$

Now the current $J_{S,e}(z,t)$ is given by (5.21) and is non-zero and independent of the electric field. It is should be obvious from the above equation that we can find an $E(z,t)$ so that the right hand side of (5.23) is negative. This result is in direct contradiction to (5.18). Therefore the theory is not mathematically consistent. This confirms the results of previous research [3][4].

Since this result is somewhat unexpected let review how we got to this point. In a mathematical theory there are elements called postulates that are assumed to be true without proof. These postulates are used to prove additional elements of the theory which are called theorems. Now what does it imply when inconsistencies appear in the theory? It implies that the postulates are not consistent.

With this in mind what are the postulates that we have introduced. The first postulate is Eq. (3.1) along with the definition of the Hamiltonian (3.2). Of course we also assume the usual mathematical apparatus of calculus and algebra. From this initial assumption the results of Section 3 follow. These are that the Heisenberg picture is gauge invariant and that there is no lower bound to the free field energy. The next set of postulates are introduced in Section 4. These are the anti-commutation relations (4.1), along with the Schrödinger equation (4.2), and the definition of $\hat{H}$ per Eq. (4.3). From these assumptions we develop the Schrödinger picture and show that it is equivalent to the Heisenberg picture. The next postulate is the form of the Schrödinger picture field operator as defined by Eq. (5.1). The field operator is defined so that it is consistent with the anti-commutation relationships. So far we have a mathematically consistent theory. From these postulates we have proven that there is no lower bound to the free field energy in both pictures.

The next key postulate is the definition of the vacuum state per equation (5.8). From this definition it is readily shown that the vacuum state is a lower bound to the free field energy. Therefore it is at the point that an inconsistency is introduced. It will be shown later in the discussion that we can redefine the vacuum state in order to remove this inconsistency.

## 7. The Schwinger term.

We shall now discuss the effect that this inconsistency has on the problem of gauge invariance. The key focus of this discussion will be the Schwinger term. As discussed in the Introduction (see Eq. (1.12)) the Schwinger term $\langle 0|\left[\hat{\rho}_S(z'),\hat{J}_S(z)\right]|0\rangle$ must be zero in order for the vacuum current to be gauge invariance. This can also be shown more directly as follows. Consider the time derivative of the current expectation value in the Schrödinger picture,

$$\frac{\partial J_{S,e}(z,t)}{\partial t} = i\left\langle \Omega(t) \middle| \left[ \hat{H}, \hat{J}_S(z) \right] \middle| \Omega(t) \right\rangle \tag{6.1}$$

Using (4.3) and (4.19) we can show that,

$$\hat{H} = \hat{H}_0 - \int \hat{J}_S(z) \cdot A_1(z,t) dz + \int \hat{\rho}_S(z) A_0(z,t) dz \tag{6.2}$$

Use this in (6.1) to obtain,

$$\frac{\partial J_{S,e}(z,t)}{\partial t} = i\left\langle \Omega(t) \right| \begin{pmatrix} \left[ \hat{H}_0, \hat{J}_S(z) \right] - \int \left[ \hat{J}_S(z') \cdot A_1(z',t), \hat{J}_S(z) \right] dz' \\ + \int \left[ \hat{\rho}_S(z'), \hat{J}_S(z) \right] A_0(z',t) dz' \end{pmatrix} \left| \Omega(t) \right\rangle \tag{6.3}$$

Next perform the gauge transformation (1.2) to obtain,

$$\frac{\partial J_{S,e}(z,t)}{\partial t} = i\left\langle \Omega(t) \right| \begin{pmatrix} \left[ \hat{H}_0, \hat{J}_S(z) \right] - \int \left[ \hat{J}_S(z') \cdot \left( A_1(z',t) - \frac{\partial \chi(z',t)}{\partial z'} \right), \hat{J}_S(z) \right] dy \\ + \int \left[ \hat{\rho}_S(z'), \hat{J}_S(z) \right] \left( A_0(z',t) + \frac{\partial \chi(z',t)}{\partial t} \right) dz' \end{pmatrix} \left| \Omega(t) \right\rangle$$

(6.4)

The quantity $\partial J_{S,e}/\partial t$ is a physical observable and therefore, if the theory is gauge invariant, must not depend on the quantities $\chi$ or $\partial\chi/\partial t$. Now, at a particular instant of time $\partial\chi/\partial t$ can be varied in an arbitrary manner without changing the values of any of the other quantities on the right hand side of the equals sign in the above equation. Therefore for $\partial J_{S,e}/\partial t$ to be independent of $\partial\chi/\partial t$ we must have that the quantity $\left[ \hat{\rho}_S(z'), \hat{J}_S(z) \right] = 0$. However it has been shown by Schwinger [6] and by Ref. [3] that this quantity cannot be zero if the vacuum state is the state of minimum free field energy.

To verify this we will calculate the Schwinger term $\left\langle 0 \middle| \left[ \hat{\rho}_S(z'), \hat{J}_S(z) \right] \middle| 0 \right\rangle$. If $\left[ \hat{\rho}_S(z'), \hat{J}_S(z) \right] = 0$ then the Schwinger term will be zero. Also if the Schwinger term is

not zero then $\left[\hat{\rho}_S\left(z'\right),\hat{J}_S\left(z\right)\right]$ is not zero. In the following discussion it will be shown that the Schwinger term is not zero. Using (5.1) we can show that the Schrödinger current operator can be written as,

$$\hat{J}_S\left(z\right)=q\hat{\psi}_0^\dagger\left(z\right)\sigma_3\hat{\psi}_0\left(z\right)=\hat{J}_{+1}\left(z\right)+\hat{J}_{-1}\left(z\right) \tag{6.5}$$

where,

$$\hat{J}_{\pm 1}\left(z\right)=q\hat{\psi}_{0,\pm 1}^\dagger\left(z\right)\sigma_3\hat{\psi}_{0,\pm 1}\left(z\right) \tag{6.6}$$

$$\hat{\rho}_S\left(z\right)=q\hat{\psi}_0^\dagger\left(z\right)\psi_0\left(z\right)=\hat{\rho}_{+1}\left(z\right)+\hat{\rho}_{-1}\left(z\right) \tag{6.7}$$

where

$$\hat{\rho}_{\pm 1}\left(z\right)=q\hat{\psi}_{0,\pm 1}^\dagger\left(z\right)\hat{\psi}_{0,\pm 1}\left(z\right) \tag{6.8}$$

We can easily show that $\left\langle 0\middle|\left[\hat{\rho}_{+1}\left(z'\right),\hat{J}_{-1}\left(z\right)\right]\middle|0\right\rangle=\left\langle 0\middle|\left[\hat{\rho}_{-1}\left(z'\right),\hat{J}_{+1}\left(z\right)\right]\middle|0\right\rangle=0$. Use this to obtain,

$$\left\langle 0\middle|\left[\hat{\rho}_S\left(z'\right),\hat{J}_S\left(z\right)\right]\middle|0\right\rangle=\left\langle 0\middle|\left[\hat{\rho}_{+1}\left(z'\right),\hat{J}_{+1}\left(z\right)\right]\middle|0\right\rangle+\left\langle 0\middle|\left[\hat{\rho}_{-1}\left(z'\right),\hat{J}_{-1}\left(z\right)\right]\middle|0\right\rangle \tag{6.9}$$

First evaluate $\left\langle 0\middle|\left[\hat{\rho}_{+1}\left(z'\right),\hat{J}_{+1}\left(z\right)\right]\middle|0\right\rangle$,

$$\left\langle 0\middle|\left[\hat{\rho}_{+1}\left(z'\right),\hat{J}_{+1}\left(z\right)\right]\middle|0\right\rangle=\left\langle 0\middle|\left[\hat{\rho}_{+1}\left(z'\right)\hat{J}_{+1}\left(z\right)\right]\middle|0\right\rangle-c.c. \tag{6.10}$$

Use (5.2) and (5.8) in the above to obtain,

$$\begin{aligned}\left\langle 0\middle|\left[\hat{\rho}_{+1}\left(z'\right)\hat{J}_{+1}\left(z\right)\right]\middle|0\right\rangle=\left\langle 0\right|&\left(\sum_{q>0}\hat{d}_{q,+1}\phi_{-q,+1}^\dagger\left(z'\right)\right)\left(\sum_{p>0}\begin{pmatrix}\hat{b}_{p,+1}\phi_{p,+1}\left(z'\right)\\+\hat{d}_{p,+1}^\dagger\phi_{-p,+1}\left(z'\right)\end{pmatrix}\right)\\&\text{times}\left(\sum_{p'>0}\begin{pmatrix}\hat{b}_{p',+1}^\dagger\phi_{p',+1}^\dagger\left(z\right)\\+\hat{d}_{p',+1}\phi_{-p',+1}^\dagger\left(z\right)\end{pmatrix}\right)\sigma_3\left(\sum_{q'>0}\hat{d}_{q',+1}^\dagger\phi_{-q',+1}\left(z\right)\right)\left|0\right\rangle\end{aligned} \tag{6.11}$$

Use (5.4) and (5.5) in the above to obtain,

$$\left\langle 0\middle|\left[\hat{\rho}_{+1}\left(z'\right)\hat{J}_{+1}\left(z\right)\right]\middle|0\right\rangle=\frac{1}{L^2}\left(\sum_{q>0,p>0}e^{iq(z'-z)}e^{ip(z'-z)}+\left(\sum_{p>0}1\right)^2\right) \tag{6.12}$$

Use this in (6.10) to yield,

$$\left\langle 0\middle|\left[\hat{\rho}_{+1}\left(z'\right),\hat{J}_{+1}\left(z\right)\right]\middle|0\right\rangle=\frac{2i}{L^2}\sum_{q>0,p>0}\sin\left(\left(p+q\right)\left(z'-z\right)\right) \tag{6.13}$$

Similarly it can be shown that,

$$\langle 0|\left[\hat{\rho}_{-1}(z'),\hat{J}_{-1}(z)\right]|0\rangle = -\frac{2i}{L^2}\sum_{q<0,p<0}\sin\left((p+q)(z'-z)\right) = \frac{2i}{L^2}\sum_{q>0,p>0}\sin\left((p+q)(z'-z)\right)$$

(6.14)

Use the above results in (6.9) to obtain,

$$\langle 0|\left[\hat{\rho}_S(z'),\hat{J}_S(z)\right]|0\rangle = \frac{4i}{L^2}\sum_{q>0,p>0}\sin\left((p+q)(z'-z)\right) \tag{6.15}$$

Now we want to determine if this quantity is zero for all $(z'-z)$. If it is zero then its first derivative with respect to $z'$ should be zero. Take this derivative to obtain,

$$\frac{\partial\langle 0|\left[\hat{\rho}_S(z'),\hat{J}_S(z)\right]|0\rangle}{\partial z'} = \frac{4i}{L^2}\sum_{q>0,p>0}(p+q)\cos\left((p+q)(z'-z)\right) \tag{6.16}$$

Next let $z'=z$ to yield,

$$\frac{\partial\langle 0|\left[\hat{\rho}_S(z'),\hat{J}_S(z)\right]|0\rangle}{\partial z'} \underset{z'=z}{=} \frac{4i}{L^2}\sum_{q>0,p>0}(p+q) \neq 0 \tag{6.17}$$

Therefore the Schwinger term $\langle 0|\left[\hat{\rho}_S(z'),\hat{J}_S(z)\right]|0\rangle$ is non-zero.

## <u>8. Redefining the vacuum state.</u>

We have shown that there is no lower bound to the free field energy in both the Heisenberg and Schrödinger pictures. However we have defined the vacuum state so that there is a lower bound to the free field energy. This makes the theory mathematically inconsistent. We have also shown that this definition of the vacuum state means that the Schwinger term must be non-zero which destroys the gauge invariance of the theory. The next logical step is to attempt to define the vacuum so that it is not the state of minimum free field energy. This has already been done in Ref. [4] where it was shown that when we redefine the vacuum state so that this requirement is met we obtain a gauge invariant theory. Here we take an approach similar to that of [4].

Define the state $|0_R\rangle$ according to,

$$\hat{b}_{p,s}|0_R\rangle = 0 \text{ for all } p\text{; } \hat{d}_{p,s}|0_R\rangle = 0 \text{ for } |p|<R\text{; } \hat{d}^{\dagger}_{p,s}|0_R\rangle = 0 \text{ for } |p|>R \tag{7.1}$$

where $R$ is a positive number. Now as $R\to\infty$ we can see that in a certain sense $|0_R\rangle$ can be said to approach $|0\rangle$ (compare (7.1) with (5.8) as $R\to\infty$). Therefore we take $\left|0_R\right\rangle$ to be the vacuum state where $R$ is an arbitrarily large positive number. Using (5.7) and (7.1) we obtain,

$$\hat{H}_0\left|0_R\right\rangle = \left(-\sum_s \sum_{p=-R}^{p=+R} |p|\right)\left|0_R\right\rangle = \varepsilon\left(\left|0_R\right\rangle\right)\left|0_R\right\rangle \tag{7.2}$$

where $\varepsilon\left(\left|0_R\right\rangle\right)$ is the energy eigenvalue of the state $\left|0_R\right\rangle$ and is defined by the above expression. New states are formed by acting on the state $\left|0_R\right\rangle$ with (1) the operator $\hat{b}^\dagger_{p,s}$ for all $p$; (2) the operator $\hat{d}^\dagger_{p,s}$ for $|p| < R$; (3) and $\hat{d}_{p,s}$ for $|p| > R$. When we act on $\left|0_R\right\rangle$ with the raising operators a new state is produced with free field energy greater than $\left|0_R\right\rangle$. If we act on $\left|0_R\right\rangle$ with the destruction operator $\hat{d}_{q,s}$ with $|q| > R$ then the free field energy of the state $\hat{d}_{q,s}\left|0_R\right\rangle$ can easily be shown to be,

$$\left\langle 0_R\right|\hat{d}^\dagger_{q,s}\hat{H}_0\hat{d}_{q,s}\left|0_R\right\rangle = \varepsilon\left(\left|0_R\right\rangle\right) - |q| \tag{7.3}$$

Therefore the free field energy of this state is less than that of the state $\left|0_R\right\rangle$. Therefore if $\left|0_R\right\rangle$ is used as the vacuum state instead of $\left|0\right\rangle$ then there will exist states with less free field energy than that of the vacuum state $\left|0_R\right\rangle$. As previously discussed this is required for a mathematically consistent theory.

Now we shall calculate the Schwinger term using $\left|0_R\right\rangle$ as the vacuum state. In this case we evaluate $\left\langle 0_R\right|\left[\hat{\rho}_S(z'), \hat{J}_S(z)\right]\left|0_R\right\rangle$. Based on the discussion in Section 6 we have,

$$\left\langle 0_R\right|\left[\hat{\rho}_S(z'), \hat{J}_S(z)\right]\left|0_R\right\rangle = \left\langle 0_R\right|\left[\hat{\rho}_{+1}(z'), \hat{J}_{+1}(z)\right]\left|0_R\right\rangle + \left\langle 0_R\right|\left[\hat{\rho}_{-1}(z'), \hat{J}_{-1}(z)\right]\left|0_R\right\rangle \tag{7.4}$$

First evaluate,

$$\left\langle 0_R\right|\left[\hat{\rho}_{+1}(z'), \hat{J}_{+1}(z)\right]\left|0_R\right\rangle = \left\langle 0_R\right|\left[\hat{\rho}_{+1}(z')\hat{J}_{+1}(z)\right]\left|0_R\right\rangle - c.c. \tag{7.5}$$

From the previous discussion we obtain,

$$\langle 0_R | \left[ \hat{\rho}_{+1}(z') \hat{J}_{+1}(z) \right] | 0_R \rangle = \langle 0_R | \left( \sum_{q'=0}^{R} \hat{d}_{q',+1} \phi^{\dagger}_{-q',+1}(z') \right) \left( \sum_{p'=0}^{\infty} \begin{pmatrix} \hat{b}_{p',+1} \phi_{p',+1}(z') \\ +\hat{d}^{\dagger}_{p',+1} \phi_{-p',+1}(z') \end{pmatrix} \right)$$
$$\text{times} \left( \sum_{p=0}^{\infty} \begin{pmatrix} \hat{b}^{\dagger}_{p,+1} \phi^{\dagger}_{p,+1}(z) \\ +\hat{d}_{p,+1} \phi^{\dagger}_{-p,+1}(z) \end{pmatrix} \right) \left( \sum_{q=0}^{R} \hat{d}^{\dagger}_{q,+1} \phi_{-q,+1}(z) \right) | 0_R \rangle \tag{7.6}$$

Use (7.1) to obtain,

$$\langle 0_R | \left[ \hat{\rho}_{+1}(z') \hat{J}_{+1}(z) \right] | 0_R \rangle = \frac{1}{L^2} \left( \sum_{q=0}^{R} \sum_{p=0}^{\infty} e^{iq(z'-z)} e^{ip(z'-z)} + \sum_{q=0}^{R} \sum_{p=R}^{\infty} e^{iq(z'-z)} e^{ip(z-z')} + \left( \sum_{p=0}^{R} 1 \right)^2 \right) \tag{7.7}$$

Use this in (7.5) to yield,

$$\langle 0_R | \left[ \hat{\rho}_{+1}(z'), \hat{J}_{+1}(z) \right] | 0_R \rangle = \frac{1}{L^2} \sum_{q=0}^{R} \sum_{p=0}^{\infty} \left( e^{iq(z'-z)} e^{ip(z'-z)} - c.c. \right) + \frac{1}{L^2} \sum_{q=0}^{R} \sum_{p=R}^{\infty} \left( e^{iq(z'-z)} e^{ip(z-z')} - c.c. \right) \tag{7.8}$$

The quantity $\frac{1}{L^2} \sum_{q=0}^{R} \sum_{p=0}^{R} \left( e^{iq(z'-z)} e^{ip(z-z')} - c.c. \right)$ equals zero so we can add this to the above equation. When this is done we obtain,

$$\langle 0_R | \left[ \hat{\rho}_{+1}(z'), \hat{J}_{+1}(z) \right] | 0_R \rangle = \frac{1}{L^2} \sum_{q=0}^{R} \sum_{p=-\infty}^{\infty} \left( e^{iq(z'-z)} e^{ip(z'-z)} - c.c. \right) = \left( \frac{1}{L^2} \sum_{q=0}^{R} e^{iq(z'-z)} \sum_{p=-\infty}^{\infty} e^{ip(z'-z)} \right) - c.c. \tag{7.9}$$

Use $\sum_{p=-\infty}^{\infty} e^{ip(z'-z)} = \delta(z'-z)$ in the above to obtain,

$$\langle 0_R | \left[ \hat{\rho}_{+1}(z'), \hat{J}_{+1}(z) \right] | 0_R \rangle == \left( \frac{1}{L^2} \sum_{q=0}^{R} e^{iq(z'-z)} \delta(z'-z) \right) - c.c. = \left( \frac{1}{L^2} \sum_{q=0}^{R} \delta(z'-z) \right) - c.c = 0 \tag{7.10}$$

where we have used $e^{iq(z'-z)} \delta(z'-z) = \delta(z'-z)$ which follow from the definition if the dirac delta function. Similarly we can show that $\langle 0_R | \left[ \hat{\rho}_{-1}(z'), \hat{J}_{-1}(z) \right] | 0_R \rangle = 0$. Therefore we obtain,

$$\langle 0_R | \left[ \hat{\rho}_S(z'), \hat{J}_S(z) \right] | 0_R \rangle = 0 \tag{7.11}$$

Therefore we have shown that when we use $|0_R\rangle$ as the vacuum state then the Schwinger term is zero. This means that the theory will be gauge invariant. This result is consistent with the observation by Schwinger[6] that in order for the Schwinger term to be zero that must exist states whose free field energy is less than that of the vacuum state.

One possible objection to using the state $|0_R\rangle$ is that is will somehow decay into some lower energy state due to perturbations of the electric field. We can easily show that this is not the case by referring to Eq. (3.28) where we calculated the change in the free field energy due to an applied electric field. Recalling that the Schrödinger and Heisenberg pictures are equivalent we have that the change in the free field energy of the state $|0_R\rangle$ due to an applied electric field is,

$$\Delta\xi_0\left(|0_R\rangle\right)=\int_0^{t_f}dt\int dz\left\langle 0_R\right|\hat{\psi}_0^\dagger(z,t)\sigma_3\hat{\psi}_0(z,t)\left|0_R\right\rangle E(z,t) \tag{7.12}$$

Now $\left\langle 0_R\right|\hat{\psi}_0^\dagger(z,t)\sigma_3\hat{\psi}_0(z,t)\left|0_R\right\rangle$ is the current expectation value of the unperturbed vacuum state which is zero. Therefore $\Delta\xi_0\left(|0_R\rangle\right)$ is zero for any arbitrary electric field. Therefore a "dramatic" event such as the vacuum decaying into some lower energy state is not going to occur.

## 9. Conclusion.

We have examined a simple field theory consisting of quantized fermion field with zero mass fermions in a classical electric field in 1-1D space-time. The theory is simple enough that we can obtain an exact solution in both the Heisenberg and Schrödinger pictures. We have shown that both pictures are equivalent and that there is no lower bound to the free field energy. This requires that the vacuum state be properly defined. If the vacuum state is defined the traditional way per Eq. (5.8) then the theory will be mathematically inconsistent. That is because $|0\rangle$ will be the minimum state of free field energy.

We have also calculated the Schwinger term and shown that it is non-zero if $|0\rangle$ is used for the vacuum state. As was discussed in Section 7 in order for quantum field theory to be gauge invariant the Schwinger term must be zero. Therefore the use of $|0\rangle$ as the vacuum state destroys the gauge invariance of the theory.

This raises the obvious question as to whether or not the vacuum state can be redefined so that it is consistent with the requirements of gauge invariance. One possible way to do this was discussed in Section 7. Here the state $|0_R\rangle$ was defined so that it was not a lower bound to the free field energy. This is required for a mathematically consistent theory. When this redefined vacuum state $|0_R\rangle$ is used the Schwinger term was shown to be zero. Therefore the theory will be gauge invariant. A possible objection to using $|0_R\rangle$ is that it will decay into a lower energy state. However we have shown that this is not the case. An interaction with an electric field will not change the free field energy of the state $|0_R\rangle$. This suggests that in order to obtain a mathematically consistent field theory which is gauge invariant we must considering redefining the vacuum state along the lines suggested in Section 7 of this paper.

## References

1. J. Schwinger, Phys. Rev. **81**, 664 (1951).
2. W. Greiner, B. Muller, and J. Rafelski, "Quantum Electrodynamics of Strong Fields", Springer-Verlag, Berlin (1985).
3. D. Solomon, Phys. Scr., **76** (2007), 64. See also arXiv.0706.2830
4. D. Solomon, Can. J. Phys. **76**(1998), 111. See also arXiv:quant-ph/9905021.
5. W. Heitler. The quantum theory of radiation. Dover Publications, Inc., New York (1954).
6. J. Schwinger, Phys. Rev. Lett., **3**, 296 (1959).